\documentclass[12pt]{article}
\newcommand{\be}{\begin{equation}}
\newcommand{\ee}{\end{equation}}
\newcommand{\beas}{\begin{eqnarray*}}
\newcommand{\eeas}{\end{eqnarray*}}
\newcommand{\bea}{\begin{eqnarray}}
\newcommand{\eea}{\end{eqnarray}}
\newcommand{\ba}{\begin{array}}
\newcommand{\ea}{\end{array}}
\newcommand{\nn}{\nonumber}

\newcommand{\de}{\delta}
\newcommand{\la}{\lambda}
\newcommand{\si}{\sigma} 
\newcommand{\n}{\nabla}

\begin{document}
\title{
{\bf
Scalar graviton and the modified black holes
}}
\author{Yu.\ F.\ Pirogov
\\
\it 
Institute for High Energy Physics,  Protvino, Russia
}
\date{}
\maketitle
\abstract{\noindent 
Under the explicit violation of the general covariance to the
unimodular one, the effect of the  emerging scalar graviton on the
static spherically symmetric metrics is
studied. The main results are three-fold. First, there appears the
two-parametric family of such metrics, instead of the one-parametric
black-hole family in General Relativity (GR). Second, 
there may exist the one-parametric subfamily describing a pure
gravitational object, the ``graviball'', missing in GR. Third,
in a simplifying assumption, all the metrics possess the correct
Newton's limit as in GR.
}

\section{Introduction}

In paper~\cite{Pir1}, we proposed the metric effective field theory
with the explicit violation of the general covariance (GC) to the
residual unimodular covariance (UC). Due to such a violation, the
resulting modification of General Relativity (GR) describes the
massive scalar graviton as a part of the metric, in addition to the
massless tensor graviton as in GR. Associated with the scalar
graviton, there appears a dimensionful parameter which is a~priori
arbitrary. The  scalar graviton was proposed as a source of the
dark matter and the dark energy of the gravitational
origin.\footnote{For a brief exposition of the respective
topics, see~\cite{Pir2}.}  In a subsequent paper~\cite{Pir3},
this concept was applied to studying the evolution of the isotropic
homogeneous Universe. It was found there that to treat the
scalar graviton as the dominant source of the cold dark matter,  the
aforementioned dimensionful parameter should be large, of the order of
the Planck mass. It follows thereof that such a parameter could
strongly invalidate the Newton's limit for the metric of the
gravitating center. This question is studied in the given paper. 

In Section 2, the theory of gravity with UC  and the scalar graviton
is briefly reviewed to be used in what follows. In Section 3,  the
static spherically symmetric metric of a
point-like body, together with  the surrounding distribution of the
scalar gravitons, is investigated. The main results are three-fold.
First, there exists the two-parametric family of the respective
metrics instead of the one-parametric  GR family, the black holes. 
Second, among the metrics, there may exist those for a peculiar
object, the ``graviball''. Third, in a simplifying
assumption, all the metrics possess the correct Newton's limit without
fine tuning, making thus the proposed GR modification as  robust
in the Newtonian approximation as GR~itself.  
What remains to be done is indicated in the Conclusion.

\section{Scalar graviton}

\paragraph{Gravity action}

Let us remind in brief the metric effective field theory
with~UC~\cite{Pir1}. In addition to
the massless tensor graviton, such a theory
describes the massive scalar  one as a part of the metric field. In
the vacuum, the gravity action consists generically of two parts:
\be\label{S}
S= S_{ g}+S_{ s}.
\ee
The first, generally covariant part of the action, responsible
for the massless tensor graviton, is  as in GR:
\be\label{GR}
S_{ g}=-\frac{1}{2}m_P^2\int R(g_{\mu\nu})
\sqrt{-g}d^4 x,
\ee
with $m_P=(8\pi G_{ N})^{-1/2}$ being the  Planck mass and 
$G_{ N}$  the Newton's constant. 
In the above, $x^\mu$, $\mu=0,\dots, 3$, are the arbitrary observer's
coordinates, $g_{\mu\nu}$ is  the metric, $g\equiv \mbox{\rm det\,}
g_{\mu\nu}$,  $R=g^{\mu\nu} R_{\mu\nu}$ is the Ricci scalar,
with $R_{\mu\nu}$  being the Ricci curvature tensor: 
\be\label{Rmunu}
R_{\mu\nu}=\partial_\la \Gamma^\la{}_{\mu\nu}- \partial_\mu
\Gamma^\la{}_{\la\nu}   +  \Gamma^\la{}_{\mu\nu}
\Gamma^\rho{}_{\rho\la}
- \Gamma^\la{}_{\mu\rho}  \Gamma^\rho{}_{\nu\la} .
\ee
Also, $\Gamma^\la{}_{\mu\nu}$ is the
Christoffel connection: 
\be
\Gamma^{\la}{}_{\mu\nu}=\frac{1}{2} g^{\la\rho}(
\partial_\mu g_{\nu\rho}+\partial_\nu g_{\mu\rho}
-\partial_\rho g_{\mu\nu}),
\ee
so that $\Gamma^\la{}_{\la\nu}=\partial_\nu \ln\sqrt{-g}$
and thus $\partial_\mu \Gamma^\la{}_{\la\nu}=\partial_\mu\partial_\nu
\ln\sqrt{-g}$.

The second part of the gravity action,  describing the massive scalar
graviton, looks in the lowest derivative order as
\be\label{Lsi}
S_{ s} =\int \Big(\frac{1}{2}\partial \si\cdot \partial \si
-V_{ s}(\si)\Big)\sqrt{-g}d^4 x,
\ee
with $\partial \si\cdot \partial \si=g^{\mu\nu}\partial_\mu\si
\partial_\nu\si$, etc. Here, $\si$ is the field of the scalar
graviton: 
\be 
\si= f_{ s}\ln\sqrt{g/\tilde g},
\ee
with $\tilde g $  being a nondynamical scalar density of the same
weight as the scalar density~$g$. Such an object  is necessary to
build the scalar out of $g$. Knowing $\tilde g $ in the observer's
coordinates, one can always rescale the latter ones locally, so that
$\tilde g=-1 $. On the other hand, the observer's coordinates being
fixed, changing $\tilde g$ by a multiplicative constant results in
shifting $\si$ by an additive constant. 
In Eq.~(\ref{Lsi}), $f_{ s}$ is a constant with the dimension of
mass. A priori, one expects $f_{ s}\leq {\cal O}(m_P)$.
$V_{ s}$ is a potential producing the mass for the scalar
graviton. By the symmetry reasons,  the potential is supposed to be
suppressed. 

Eq.~(\ref{Lsi}) violates GC, still preserving UC. In
particular, this insures that the causality is not violated. 
Such a modified GR is the bona fide  quantum field
theory, as robust theoretically as GR itself~\cite{Pir2}.
Nevertheless, there being no smooth restoration of GR at the
quantum level, the proposed GR modification is
in fact an independent theory, which is to be verified
experimentally.

\paragraph{Gravity equations}

Varying the gravity action with respect to $g^{\mu\nu}$,
$\tilde g$ being fixed, one arrives at the modified gravity
equation in the vacuum:
\be\label{eomg}
R_{\mu\nu}-\frac{1}{2} R g_{\mu\nu} = \frac{1}{m_{ P}^2}
T_{s \mu\nu},
\ee
where  
\be\label{DT}
T_{s\mu\nu}=
\partial_\mu\si \partial_\nu\si-
\frac{1}{2}\partial\si\cdot \partial\si\, g_{\mu\nu}  +
V_{ s} g_{\mu\nu}
+ f_{ s}(\n\cdot\n \si + \partial V_{ s}/\partial \si)
g_{\mu\nu}
\ee
is to be treated as the metric energy-momentum tensor of the scalar
graviton. In the above
\be
\n\cdot\n \si=\frac{1}{\sqrt {-g}}
\partial_\mu(\sqrt {-g}g^{\mu\nu} \partial_\nu\si). 
\ee
Due to the Bianchi identity 
\be
\nabla_\mu \Big(R^{\mu\nu}- \frac{1}{2} R g^{\mu\nu}\Big) =0,
\ee
the total energy-momentum of the ordinary
matter, omitted in Eq.~(\ref{S}), and the scalar graviton  satisfies
the continuity condition. For this reason, the scalar graviton can
naturally be considered as a source of the dark matter
and the dark energy of the gravitational origin. In the absence of the
ordinary matter, the continuity condition  reduces~to 
\be\label{cc}
\nabla_\mu  T_{ s}^{\mu\nu} =0.
\ee

The modified pure gravity equation~(\ref{eomg}) can otherwise be
written as
\be\label{eomg'}
R_{\mu\nu}= \frac{1}{m^2_{ P}} \Big(
T_{ s\mu\nu}-\frac{1}{2}
T_{ s} g_{\mu\nu}\Big),
\ee
where $T_{ s}\equiv g_{\mu\nu} T_{ s}^{\mu\nu}$ is as follows:
\be
T_{ s}= -
\partial\si\cdot\partial\si +4V_{ s}+4f_{ s}(\n\cdot\n \si
+\partial V_{ s}/\partial \si), 
\ee 
with $ R=-T_{ s}/m^2_{ P}$ generally nonzero. 
Explicitly, one gets 
\be\label{eomg''}
R_{\mu\nu} = \frac{1}{m_{ P}^2}\Big(
\partial_\mu \si  \partial_\nu \si - V_{ s} g_{\mu\nu}
-f_{ s}(\n\cdot\n \si+
\partial V_{ s}/\partial \si)g_{\mu\nu}\Big).
\ee

\section{Modified black holes}

\paragraph{Comoving coordinates}

Under the violation of GC, to really start
up the theory one should choose the particular coordinates where
$\tilde g$ is to be defined. 
There is a natural choice. Namely, 
according to the present-day cosmological paradigm, our Universe 
is spatially flat, fairly homogeneous, and isotropic. Thus, given a
point~$P$, the observer can adjust the coordinates in the Universe so
that the metric around the  point at the astronomical scales, much
less then the cosmological ones, may be put to the Minkowskian form, 
$g_{\mu\nu}=\eta_{\mu\nu}$.  Call these coordinates the comoving ones. 
In the comoving coordinates, the nondynamical variable ${\tilde g}$,
characterizing the Universe as a whole, is to vary at the cosmological
scales as the Universe itself. So, ${\tilde g}$ can be
treated in the given approximation as a constant, too.
After the theory starts up in the comoving coordinates,
it can be rewritten in the arbitrary observer's coordinates.

Now, let in the point $P$ there be placed the isolated point-like body
which is in the rest  relative to the close ambient.
The body is assumed to have no angular momentum and other physical
attributes, but the mass $m$. Such a body disturbs
the Universe metric in a vicinity of the point, not violating the
spherical symmetry.
The interval corresponding to the static spherically symmetric metric
around such a body can be chosen in the comoving coordinates 
most generally as follows:
\be\label{ds2}
ds^2= a(r)d t^2 -
\Big(b( r)- c(r)\Big)({\bf  n}
d {\bf x})^2
- c( r)d{\bf x}^2,
\ee
where  $ r^2= {\bf  x}^2 =\delta_{mn}  x^m   x^n $,
$m,n=1,2,3$, etc., $n^m=  x^m/ r$,  $n_m=\de_{ml} n^l$, with 
$ {\bf n}^2=1$. The functions $ a(r)$, $b(r)$, and $c(r)$ are the
dynamical metric variables to be determined through the gravity
equations. The latter ones are to be supplemented by the
asymptotic condition: $ a(r)$, $b(r)$, $c(r)\to 1$ at $r\to \infty$.

The respective metric looks like
\bea\label{metric}
g_{00}&=& a,\nn\\
 g_{mn}&=&- b n_m  n_n
- c(\de_{mn}-  n_m  n_n) ,
\eea
with  the inverse metric
\bea
 g^{00}&=&\frac{1}{ a},\nn\\
 g^{mn}&=&-\frac{1}{ b}\, n^m n^n 
-\frac{1}{ c}(\de^{mn}- n^m  n^n).
\eea
The rest of the metric elements is zero. Rotating
the spatial coordinates so that in the point $\bf
x$ there takes place ${\bf n}=(1,0,0)$, one brings
the metric in this point to the diagonal
form $(g_{\mu\nu})=\mbox{\rm diag\,} (a,- b ,-c,-c)$. Thus,
generally, the metric is  anisotropic. For the isotropy,
there should be fulfilled $c=b$. 
In general, one has  $ g= - a b c^2$ and thus 
\be\label{sir'}
 \si(r)=f_{ s}\ln (\sqrt{ a b} c/ \sqrt{-\tilde g}),
\ee
with $ \si\to f_{ s}\ln(1/\sqrt{-{\tilde
g}})$ at $ r\to \infty$. 

One can change the original comoving coordinates $ x^m$ to the
polar comoving ones $(r,\theta,\varphi)$, so that the interval
becomes
\be\label{polc} 
ds^2= a d  t^2- b d   r^2-  c  r^2(d\theta^2
+\sin^2 \theta d\varphi^2),
\ee 
with a unit of length being tacitly understood where it necessary.
One has  now $g=-abc^2 r^4\sin^2\theta$, with $\si$
looking nevertheless the same as in Eq.~(\ref{sir'}) due to the
compensating transformation of~$\tilde g$. 

\paragraph{Coordinate change}

Having defined the theory in the comoving coordinates, one can
choose, not violating the spherical symmetry,  the new radial
coordinate $\hat r=\hat r(r)$, 
with the metric variables becoming as follows:
\bea\label{hat}
\hat a(\hat r)&=& a(r(\hat r)),\nn\\
\hat b(\hat r)&=&  (dr/d \hat r)^2 b(r(\hat r)),\nn\\
\hat c(\hat r)&=&  (r/\hat r)^2 c(r(\hat r)).
\eea
To preserve the asymptotic condition on the metric one should impose
the restriction $\hat r/ r\to 1$ at $r\to \infty$.
The scalar graviton distribution now looks like 
\be\label{sir,}
\hat \si(\hat r)=\si( r(\hat r))=f_{ s}\ln 
\Big(\sqrt{\hat a \hat b}\hat c\Big/\sqrt{-\hat {\tilde g}}\Big),
\ee
with
\be\label{tildeg'}
\sqrt{-\hat {\tilde g}}=  ( r/\hat r)^2 (d  r/d \hat r)
\sqrt{-{\tilde g}}.
\ee
A~priori, all the choices of $\hat r$ are equivalent under the
reversibility. Due to this,  by imposing a
restriction on $\hat c$ one can bring the metric to
the  form  most appropriate for
the particular purposes. At that, the role of the third independent
dynamical variable instead of $\hat c$ is played by  the scalar field
$\hat\si(\hat r)$. The latter becomes a kind of a hidden variable
in the absence of the direct interactions of the scalar graviton  with
the ordinary matter.

\paragraph{Isotropic form}

For example, imposing the relation $\hat c=\hat b$ by choosing $\hat
r$
through
\be
( r/\hat r )^2 c=
 (d r/d  \hat r)^2 b,
\ee
or explicitly
\be\label{riso}
\hat r= \exp \int  \sqrt{ b/ c}\,\frac{d r}{r} ,
\ee
one brings the interval to the isotropic form
\be\label{ds2i}
ds^2= \hat  a(\hat  r) d t^2 - \hat c(\hat r) d \hat {\bf x}^2,
\ee
with $\hat r^2= \hat {\bf x}^2=\de_{mn} \hat x^m \hat x^n$. 
The scalar graviton distribution is given by Eq.~(\ref{sir,})
at $\hat c= \hat b$.

\paragraph{Astronomic form}

Otherwise, imposing $\hat c=1$, so that
\be
\hat r= r \sqrt{ c(r)} , 
\ee 
one brings the interval to the  form 
\be\label{ds2'}
ds^2=\hat a(\hat r) d t^2-\hat b(\hat r) d \hat r^2- \hat
r^2(d\theta^2+\sin^2 \theta d\varphi^2).
\ee
Such a form is preferable to compare with observations. 
The scalar graviton distribution is again given by Eq.~(\ref{sir,})
at $\hat c=1$.

\paragraph{Modified Schwarzschild equations}

In what follows, we will study the static spherically
symmetric metric in the polar comoving coordinates
$(r,\theta,\varphi)$. With account for Eq.~(\ref{polc}), one gets the
metric elements  
\be
g_{00}=a, \  g_{rr}=- b,\  g_{\theta \theta}=-cr^2,\   g_{\varphi
\varphi}=-c r^2\sin^2\theta,
\ee
with  the inverse ones
\be
g^{00}=1/a, \  g^{rr}=- 1/b,\  g^{\theta \theta}=-1/(cr^2),\
g^{\varphi
\varphi}=-1/(cr^2\sin^2\theta),
\ee
the rest of the metric elements being zero. 

Designating $a'=d a/d r$, etc., one gets the nonzero elements
of the Christoffel connection as follows: 
\bea
\Gamma^0{}_{0r} =
\frac{1}{2}\frac{a'}{a} , &&\Gamma^\theta{}_{\theta
r}=\Gamma^\varphi{}_{\varphi r}
=\frac{1}{2}\frac{(cr^2)'}{cr^2}   ,\nn\\
\Gamma^r{}_{00}=\frac{1}{2}\frac{a'}{b} ,
&&\Gamma^r{}_{\theta\theta}=-\frac{1}{2}\frac{(cr^2)'}{b}, \nn\\
\Gamma^r{}_{rr}= \frac{1}{2}\frac{b'}{b},
&&\Gamma^r{}_{\varphi\varphi}= -  \frac{1}{2}\frac{(cr^2)'}{b}  \sin^2
\theta ,\nn\\
\Gamma^\varphi{}_{\varphi\theta}
=\cot\theta,&&\Gamma^\theta{}_{\varphi\varphi}=-\sin\theta\cos \theta,
\eea
with 
\bea
\Gamma^\la{}_{\la
r}&=&\frac{1}{2}\frac{a'}{a}+\frac{1}{2}\frac{b'}{b}+\frac{(cr^2)'}
{cr^2},\nn\\
\Gamma^\la{}_{\la\theta}&=&\cot\theta.
\eea
Inserting the above expressions in Eq.~(\ref{Rmunu}), one  gets the
following nonzero elements of the Ricci tensor:
\bea
R_{00}&=&\frac{1}{2}\Big(\frac{a''}{b}-\frac{1}{2}
\frac{a'^2}{ab}-\frac{1}{2}
\frac{a'b'}{b^2}\Big) +
\frac{1}{2}\frac{(cr^2)'}{cr^2}\frac{a'}{b}, \nn\\
R_{rr}&=&-\frac{1}{2}\Big(\frac{a''}{a}-\frac{1}{2}\frac{a'^2}{a^2}-
\frac{1}{2}
\frac{a'b'}{ab} \Big)+
\frac{1}{2} \frac{(cr^2)'}{cr^2}\frac{b'}{b}
+\frac{1}{2}\frac{(cr^2)'^2}{(cr^2)^2}
-\frac{(cr^2)''}{cr^2} ,\nn\\ 
R_{\theta\theta}&=&
1-\frac{1}{2}\frac{(cr^2)''}{b}-\frac{1}{4}\frac{(cr^2)'}{b}
\Big(\frac{a'}{a}-\frac{b'}{b} \Big ) ,\nn\\
R_{\varphi\varphi}&=&\sin^2\theta\, R_{\theta\theta} .
\eea
Finally, with account for Eq.~(\ref{eomg''}),
this gives the looked for system of the three equations for the three
independent variables:
\bea\label{eom}
\frac{1}{2}\frac{a''}{a}- \frac{1}{4} \frac{a'}{a} \Big(\frac{a'}{a}+
\frac{b'}{b}\Big)+\nn \\
 + \frac{b}{cr^2}-\frac{1}{2}\frac{(cr^2)''}{cr^2}+\frac{1}{4}
\frac {(cr^2)'}{cr^2} \Big(\frac{a'}{a}+\frac{b'}{b}\Big)
&=& 0, \nn\\
\frac{1}{2}\frac{(cr^2)'^2}{(cr^2)^2}-\frac{(cr^2)''}{cr^2}
+ \frac{1}{2} \frac{(cr^2)'}{cr^2}\Big(\frac{a'}{a} +\frac{b'}{b}\Big)
&=&\frac{1}{m^2_{ P}}\si'^2  , \nn\\
\frac{b}{cr^2}-\frac{1}{2}\frac{(cr^2)''}{cr^2}-\frac{1}{4}
\frac {(cr^2)'}{cr^2} \Big(\frac{a'}{a}-\frac{b'}{b}\Big) 
&=& \frac{1}{m^2_{ P} }b\Big( f_{ s}(
\n\cdot\n\si+ \partial V_{ s}/\partial\si)+ V_{ s}\Big),\ \ \ \ \ \
\eea
where 
\be
\n\cdot\n\si= -\frac{1}{b}\Big( \si''+(\ln \sqrt{a/b}cr^2)'\si'\Big),
\ee
with $\si$ given by Eq.~(\ref{sir'}). 
At $\tilde g=\mbox{\rm const}$ one has
\be\label{abc'}
\si'=f_{ s} (\ln \sqrt{ab}c)'
\ee 
and
\be\label{con}
\n\cdot\n\si= -\frac{f_{ s}}{b}\Big( (\ln\sqrt{ab}c)''
+ \frac{2}{r}(\ln \sqrt{ab}c)'
+ (\ln \sqrt{a/b}c)' (\ln \sqrt{ab}c)' \Big).
\ee

The system~(\ref{eom}) supersedes the Schwarzschild equations, valid
in GR (see later on).
Note that the first, independent of $\si$  equation  of the
system remains the same as in~GR. Under GC, this equation
holds true at any $c$. Under UC, the equation serves
to find $c$ or, otherwise, $\si$ in addition to $a$ and $b$.

\paragraph{Schwarzschild metric}

Let us recover the GR black-hole solution to be used as a reference
point. Put $\si= 0$,  restoring GC. Choose
conventionally $c=1$. Suppose also that 
$V_{ s}= 0$, neglecting thus by the cosmological term. 
Altogether,  from the last two equations of Eq.~(\ref{eom}), one gets
the Schwarzschild equations: 
\bea\label{eom''}
\frac{a'}{a} +  \frac{b'}{b} &=& 0, \nn\\
\frac{b-1}{r}-\frac{1}{2} \Big(\frac{a'}{a}-\frac{b'}{b}\Big) 
&=&  0. 
\eea
Accounting for the asymptotic condition at $r\to\infty$, one gets
$ab=1$ and $a=1- r_g/r$, with $r_g$
being an integration constant. The first equation of the
system~(\ref{eom})
can be shown to be satisfied, too. This reproduces the conventional
Schwarzschild interval in the astronomic form: 
\be\label{Schwarz}
ds^2=(1-r_g/r)dt^2-
\frac{1}{ 1-r_g/r}\,d r^2-
r^2(d \theta^2+\sin^2\theta d\varphi^2).
\ee
To insure the Newton's limit one should put 
$r_{ g} = 2G_N m$,  with $m$ being the mass of the
point-like central body. 
Choosing the new radial variable~$\hat  r$ through
\be
 r=  \hat r \Big(1+\frac{r_g}{4  \hat r }\Big)^2
\ee
and accounting for Eq.~(\ref{hat}) at $c=1$, one brings the
Schwarzschild interval to the isotropic form:
\be\label{isom}
ds^2= \bigg(\frac{1-r_{ g}/4 \hat r  }{
1+r_{ g}/4  \hat r  }\bigg)^2\,  dt^2-
(1+r_g/4 \hat r)^4  d \hat {\bf x}^2,
\ee
or
\be\label{isomr}
ds^2= \bigg(\frac{1-r_{ g}/4 \hat r  }{
1+r_{ g}/4  \hat r  }\bigg)^2\,  dt^2-
(1+r_g/4 \hat r)^4\Big(d\hat r^2+  \hat r^2 (d
\theta^2+\sin^2\theta d\varphi^2)\Big).
\ee
Otherwise, Eq.~(\ref{isomr}) can be found directly from
the last two equations of Eq.~(\ref{eom}) under $c=b$ and the
missing~r.h.s.

\paragraph{Newtonian approximation}

Now let $f_{ s}\neq 0$, but $V_{ s}$ can still be
neglected. Decompose the solutions  in the powers of~$1/r$:
\bea\label{approx}
a&=&1+\sum_{n=1}\frac{a_n}{r^n} ,\nn\\
b&=&1+ \sum_{n=1}\frac{b_n}{r^n} ,\nn\\
c&=&1+  \sum_{n=1}\frac{c_n.}{r^n}.
\eea
Substituting this decomposition into Eq.~(\ref{eom})  
note first of all that, according to equations~(\ref{abc'}) and
(\ref{con}), the r.h.s.\ of  Eq.~(\ref{eom})  appears only in the
order ${\cal O}(1/r^4)$, whereas the l.h.s.\ is nonzero already in 
${\cal O}(1/r^3)$. This means that in the leading order the modified
black-hole solutions possess asymptotically the same properties as in
GR. In particular, this insures the validity of  the Newton's limit
independent of $f_{ s}$.
More particularly, one gets the restriction
\be
a_1+b_1=0,
\ee
with $c_1$ remaining arbitrary. This reflects the absence of the  GC
violation in the gravity equations in the given
approximation. Namely, introducing the new radial coordinate
\be\label{rhatr}
\hat r= r\Big(1+\sum_{n=1}\frac{\lambda_n}{r^n}\Big)
\ee
one gets,
according to Eq.~(\ref{hat}),
\bea\label{gauge}
\hat a_1&=&a_1,\nn\\
\hat b_1&=&b_1,\nn\\
\hat c_1&=&c_1-2\lambda_1,
\eea
so that $c_1$ is indeed defined in the Newtonian approximation up to
an additive constant as in GR.

To find the solution in the post-Newtonian approximation one should 
put 
\be
b_1=-a_1=r_{ g}
\ee
and fix some $c_1$. 
In GR, this would produce $a_2$ and $b_2$ depending on $b_1$ and
$c_1$, the parameter $c_2$ being again arbitrary due to the exact
GC. Repeating this procedure in GR in all the orders, one can, in
principle, recover the Schwarzschild solution up to an arbitrary
function $c(r)$. In particular at $c=1$ and $c=b$, the proper
coefficients
$a_n$ and $b_n$
can be read off from equations (\ref{Schwarz}) and  
(\ref{isomr}), respectively.

Now under UC, due to the violation of GC
in the r.h.s.\ of  Eq.~(\ref{eom}) in the higher orders, $c_2$ is not
arbitrary but is determined together with $a_2$ and $b_2$.
Iterating this procedure, one can built the solution depending on  the
two constants $b_1$ and $c_1$. 
Thus there should exist the two-parametric
family of the modified black holes
superseding the one-parametric family of the black holes  in GR. 
Though the parameter $c_1$ gauges
according to Eq.~(\ref{gauge}), it can not be disposed at all. Namely,
decompose $\si$ as 
\be
\si=\si_0+\sum_{n=1}\frac{\si_n }{r^n},
\ee
with $\si_0=0$ without any loose of generality.
Relative to the radial
coordinate transformations Eq.~(\ref{rhatr}) one gets
\be
\hat \si_1=\si_1,
\ee
$\si_1$ being thus invariant.  It is related with  $c_1$ in the
comoving coordinates 
\be
c_1= r_{ s}
\ee
as $\si_1=f_{ s}r_{ s}$. Similar to $r_{ g}$, the parameter  $r_{ s}$
is a~priori
arbitrary and should be found through observations. The Newton's limit
for all the metrics is correctly determined by $r_g$,
independent of  $r_{ s}$.

Finally note, that there are two conceivable  marginal cases of the
static spherically symmetric metrics. First, let $r_{ g}\neq 0$,
but $r_{ s}=0$. The proper object is nothing but
the GR black hole. On the contrary, let $r_{ g}=0$.  This metric
should describe a peculiar object which may be called
the ``graviball''. For the latter,  $a$ and $ b$ vary at least as
${\cal O}(1/r^2)$  at $r\to \infty$, whereas $c$ generally as
${\cal O}(1/r)$. Nevertheless, 
by  the change of the radial coordinate the 
metric as a whole can be brought at infinity to the form varying  at
least as ${\cal O}(1/r^2)$, exhibiting thus the Newton's limit for
the massless body. At that,  $\si$ varies still as ${\cal O}(1/r)$,
being though directly unobservable.

\section{Conclusion}

The theory of gravity with UC and the scalar graviton  
admits the two-parametric variety of the modified black
holes, ranging from the ordinary black holes to the peculiar
objects, the graviballs. The theory keeps on to be uncontradicted
explicitly. The scalar graviton potential neglected, such a GR
modification is as robust in the Newtonian approximation as GR itself. 
The post-Newtonian approximations, as well as the effect of the
potential are still mandatory.

The author is grateful to 
A.~A.~Logunov for a useful discussion.

\end{document}